\documentclass[11pt, oneside]{article}   	
\usepackage{geometry}                		
\usepackage{graphicx}				
\usepackage{amssymb}
\usepackage{amsmath}
\usepackage{enumitem}
\usepackage{graphicx}
\usepackage{parskip}
\usepackage{fancyhdr}
\usepackage{xcolor}
\usepackage{diagbox}
\usepackage{placeins}
\usepackage[ruled,vlined,linesnumbered]{algorithm2e}
\usepackage[normalem]{ulem}   
\SetKwInput{kwMotivation}{Motivation}
\newcommand*{\qed}{\null\nobreak\hfill\ensuremath{\square}}
\newcommand{\NP}{\mathcal N \mathcal P}
\newcommand{\coNP}{\text{co}-\mathcal N \mathcal P}

\newtheorem{lemma}{Lemma}
\newtheorem{definition}{Definition}

\SetKwInOut{Input}{Input}

\title{Simple Stochastic Stopping Games:\\ A Generator and Benchmark Library}
\author{Avi Rudich$^{*2}$, Isaac Rudich$^{*1}$, Rachel Rue$^{*2}$}
\date{\small $^*$ All authors contributed equally. \\
$^1$ Polytechnique Montréal, Montreal, Canada \\
$^2$ Unaffiliated
\\
}

\setlist{itemsep=3pt,parsep=0pt, topsep=0pt, partopsep=0pt}

\begin{document}
\maketitle

\begin{abstract}
Simple Stochastic Games (SSGs) were introduced by Anne Condon in 1990, as the simplest version of Stochastic Games for which there is no known polynomial-time algorithm \cite{condon1992complexity}. Condon showed that Stochastic Games are polynomial-time reducible to SSGs, which in turn are polynomial-time reducible to Stopping Games.  SSGs are games where all decisions are binary and every move has a random outcome with a known probability distribution.  Stopping Games are SSGs that are guaranteed to terminate. There are many algorithms for SSGs, most of which are fast in practice, but they all lack theoretical guarantees for polynomial-time convergence.  The pursuit of a polynomial-time algorithm for SSGs is an active area of research. This paper is intended to support such research by making it easier to study the graphical structure of SSGs. Our contributions are: (1) a generating algorithm for Stopping Games, (2) a proof that the algorithm can generate any game, (3) a list of additional polynomial-time reductions that can be made to Stopping Games, (4) an open source generator for generating fully reduced instances of Stopping Games that comes with instructions and is fully documented, (5) a benchmark set of such instances, (6) and an analysis of how two main algorithm types perform on our benchmark set.

\end{abstract}

\section{Introduction}
\label{sec:intro}

Stochastic Games were first introduced in 1953 to model situations where strategic decisions are made under conditions of uncertainty. A strategy is the set of decisions each respective player makes about how to move when the game is at a position they control, and the outcome of each move is governed by a known probability distribution. \cite{shapley1953stochastic}. There exist optimal strategies that are both global (all moves are chosen in advance of the game) and deterministic (exactly one move is chosen for each game position)  \cite{condon1992complexity}. 

There is no known polynomial-time algorithm for general stochastic games. In 1990, Condon defined Simple Stochastic Games (SSGs), gave a polynomial-time reduction of Stochastic Games to SSGs, and proved that SSGs are in $\NP\:\: \cap$ co-$\NP$ \cite{condon1992complexity}. In an SSG, there are two players, and all of the decisions made by the players are binary. Simple Stochastic Games have become a focus of research because they are the simplest version of stochastic games for which there is no known polynomial-time algorithm. They are also among the few combinatorial problems in $\NP \:\: \cap$ co-$\NP$ not known to be in $P$ \cite{condon1992complexity}. 

SSGs are played on a directed graph composed of four types of nodes: max, min, average, and terminal. The two terminal nodes are labeled 1 and 0, and have out-degree 0. All other nodes have out-degree 2.  At max nodes, the max player determines which arc to take; at min nodes, the min player determines which arc to take; and at average nodes, each arc is taken with probability $\frac{1}{2}$.  The goal of the max player is to maximize the probability that a random walk on the graph from a given start node reaches terminal-1, and the goal of the min player is to minimize that probability.   The value of a node given a pair of max/min strategies is the probability that a random walk starting from that node reaches terminal-1. A stable assignment of values is any assignment compatible with mutually optimal max/min strategies.

While it is an open question whether SSGs admit a polynomial-time algorithm, many algorithms are fast in practice \cite{condon1990algorithms,GSIA2021,KRETINSKY2022,klingler2023empirical}. These algorithms generally fall into two categories: value improvement and strategy improvement.  Both types of algorithms start from some initial position and make improvements by updating either node values or player strategies until the constraints that define an SSG are satisfied. Another approach is the permutation-improvement algorithm introduced in \cite{permutation_improvement}. The algorithm starts from an initial permutation of stochastic (generalized average) nodes where a permutation represents an ordering of values, it computes the optimal strategies for that permutation, and then it updates the permutation until optimal strategies are found.

Despite considerable work in refining existing approaches, there remains a large gap between the best known lower and upper bounds.  Halman  \cite{halman2007simple} gave an upper bound of $e^{O(\sqrt{n\log n})}$ for LP-type problems and showed that SSGs can be formulated as an LP-type problem. A lower bound can be demonstrated by constructing an example that forces the Hoffman-Karp strategy improvement algorithm, which solves an LP in each iteration, to go through a linear number of iterations \cite{klingler2023empirical}. To our knowledge, no one has been able to produce an example that requires more than a linear number of iterations.

Condon made a further simplification of SSGs by proving that {SSGs} are polynomial-time reducible to Stopping Games, which terminate with probability 1 and have a unique stable value vector \cite{condon1992complexity}. Additionally, it is possible to make several other simple reductions to remove all subgraphs that are solvable in polynomial time. Improved algorithms have been proposed that leverage the graphical structure of SSGs \cite{permutation_improvement,GSIA2021,KRETINSKY2022,condon1990algorithms,dai2009new,ibsen2012solving,auger2019solving,tripathi2011strategy}. Further exploration of the structure might produce a better understanding of why it is hard to prove polynomial-time convergence, or why it is challenging to produce hard instances.  For that purpose, it is useful to have a benchmark set of SSGs that are as simple as possible without losing complexity.

Previous research on the behavior of different types of algorithms for solving SSGs used either small or non-stopping instances of SSGs \cite{KRETINSKY2022,klingler2023empirical}. In this work, we generate a set of large stopping games. The contributions of this paper are as follows: (1) We describe an algorithm for generating Stopping Game instances. (2) We prove that our algorithm is capable of generating any Stopping Game. (3) We describe several other simple polynomial-time reductions performed by the generator. (4) We provide an open source implementation of our generator algorithm that comes with instructions and is fully documented. (5) We provide a benchmark set of fully reduced problems with games ranging from $32$ nodes to $4096$ nodes. (6) We study the behavior of Hoffman-Karp (a strategy improvement algorithm) and a permutation improvement algorithm on the problems in our benchmark set. 

\subsection{Definitions}
\label{sec:intro:def}
A \emph{Simple Stochastic Game} (SSG) is a directed graph $G$ with four types of nodes: max, min, average, and terminal nodes.  Each max, min, and average node $v$ has exactly two out-arcs. There are two terminal nodes, terminal-1 and terminal-0, with no out-arcs. By convention, the nodes of $G$ are numbered $1, \dots, n$,  with the terminals 0 and 1 numbered $n-1$ and $n$, respectively. Each node $i\in G$ is assigned a value $v_i \in [0,1]$, and the values of the terminal nodes are always set to $0$ and $1$ ($v_{n-1} =0$, $v_{n}=1$).

A \emph{stable assignment} to $G$ is an assignment of values to nodes such that:
\begin{align*}
&v_{i} = \max {( v_{j}, v_{k}) }   &&\text{for each max node $i$ with children $ j,k$}   \\
&v_{i} = \min {( v_{j}, v_{k}) }   &&\text{for each min node $i$ with children $j,k$}   \\
&v_{i} = \frac {( v_{j} + v_{k}) } {2}   &&\text{for each average node $i$ with children $j,k$}
\end{align*}

We say that nodes whose values satisfy the equations above are \textit{satisfied}.

A \emph{strategy} $\sigma$ for the max player is a set of arcs consisting of one arc $(i,j)$ from each max node $i$. Equivalently, a strategy can be represented by the \emph{strategy subgraph} $G_{\sigma}$ produced from $G$ by removing every max node out-arc not contained in $\sigma$. Similarly, a strategy $\tau$ for the min player is a set of arcs consisting of one arc $(i,j)$ from each min node $i$.  The strategy subgraph $G_{\sigma,\tau}$ is the subgraph of $G$ induced by the max/min strategy pair $(\sigma,\tau)$.   

Given a pair of strategies $(\sigma,\tau)$, the \emph{solution} value of each node is the probability that a random walk starting at that node will reach terminal-1 by following the unique out-arc from each max and min node in $G_{\sigma,\tau}$, and by randomly following one of the two out-arcs from each average node with equal probability \cite{condon1992complexity}. The solution to $G_{\sigma,\tau}$ can be found by setting the value of all nodes with no path to a terminal to zero, and then solving the system of equations below for the remaining node values \cite{condon1992complexity}. 
\begin{align*}
&v_{i} = v_{j}  &&\text{for each max node $i$ with children $j,k$, where $ j \in \sigma$}   \\
&v_{i} = v_{j}  &&\text{for each min node $i$ with children $ j,k$, where $j \in \tau$}   \\
&v_{i} = \frac {( v_{j} + v_{k}) } {2}   &&\text{for each average node $i$ with children $j,k$} 
\end{align*}

The goal of the max player is to maximize the probability that a random walk will end at terminal-1 from any given start node, and the goal of the min player is to minimize that probability. A max strategy $\sigma$ is \emph{optimal} with respect to a min strategy $\tau$ if all max nodes are satisfied in $G$ by the solution values for $G_{\sigma,\tau}$, and similarly for the optimality of a min strategy $\tau$ with respect to a max strategy $\sigma$. It was shown in \cite{condon1992complexity} that there always exists at least one mutually optimal pair of strategies $(\sigma ^{\ast},\tau ^{\ast})$, and that the vector of solution values is the same for every pair of mutually optimal strategies.  Thus, the algorithmic problem for the max player is to find a strategy that maximizes the solution value of every node if the min player plays optimally.  

The associated decision problem for SSGs is: Given a simple stochastic game $G$ and a node $v$, does there exist a max strategy $\sigma$ such that for any min strategy $\tau$, a token starting at node $v$ in $G_{\sigma,\tau}$ arrives at terminal-1 with probability at least $\frac{1}{2}$? The decision problem is known to be in $\NP \cap \coNP$ \cite{condon1992complexity}.

A \emph{Stopping Game} is a simple stochastic game $G$ with the property that for every pair of strategies $(\sigma,\tau)$ and every node $v$ in $G$, there is a path from $v$ to a terminal in $G_{\sigma,\tau}$.  Equivalently, a Stopping Game is an SSG where, for any pair of strategies, a random walk starting from any node will reach a terminal with probability 1.

Anne Condon proved in \cite{condon1992complexity}:
\begin{enumerate}
\item A Stopping Game has exactly one stable assignment, which is the solution to the game played with optimal max/min strategies. 
\item For any SSG $G$, it is possible to construct a Stopping Game $G'$ such that a solution to $G$ can be recovered in polynomial time from the unique solution to $G'$. In addition, if the max player has a strategy in $G'$ such that the start node has a value $\geq \frac{1}{2}$, then the max player also has a winning  strategy in $G$.
\end{enumerate}

It follows that if Stopping Games are in $\mathcal P$, then Simple Stochastic Games (and Stochastic Games) are in $\mathcal P$. For a Stopping Game $G$, the algorithmic problem is to find the unique stable assignment of values to nodes in $G$.

\section{Stopping Game Generator}
In this section, we provide an algorithm that generates Stopping Games, with parameters for the number of max, min and average nodes. We prove that it generates only Stopping Games, and that, given fixed parameters, every Stopping Game has a non-zero probability of being generated.  In Section \ref{sec:reductions}, we describe simple subgraphs that can be solved quickly and give a modified version of the generator that produces Stopping Games with no such subgraphs.

\subsection{Bad Subgraphs}
A simple stochastic game is a Stopping Game if and only if it does not contain certain subgraphs, which we call \emph{bad subgraphs}.   The generator outputs simple stochastic games that are guaranteed not to contain any bad subgraphs.
\begin{definition}
Let $G$ be a simple stochastic game. A subgraph $S$ of $G$ is a \textit{bad subgraph} if:
\begin{enumerate} 
\item \label{def: bad subgraph start} Every max node in $S$ has at least one arc pointing into $S$.
\item Every min node in $S$ has at least one arc pointing into $S$.
\item Every average node in $S$ has both arcs pointing into $S$.
\item \label{def: bad subgraph end}$S$ contains no terminal nodes.
\item  $S$ is strongly connected (any node in $S$ is reachable from any other node in $S$).
\end{enumerate}  
\end{definition}
\begin{lemma}\label{lemma: bad subgraph SCC}
Let $G$ be a simple stochastic game. If $S$ is a subgraph of $G$ satisfying criteria \ref{def: bad subgraph start}-\ref{def: bad subgraph end} in the definition of a bad subgraph, then $S$ contains a bad subgraph.
\end{lemma}
Proof: Let $G$ be a simple stochastic game, and let $S$ be a subgraph of $G$ satisfying criteria \ref{def: bad subgraph start}-\ref{def: bad subgraph end} in the definition of a bad subgraph. Let $(\sigma,\tau)$ be a pair of max/min strategies such that the single arc from each max and min node in $S_{\sigma,\tau}$ points back into $S_{\sigma,\tau}$.  $S_{\sigma,\tau}$ also satisfies criteria \ref{def: bad subgraph start}-\ref{def: bad subgraph end}. Any path in $G_{\sigma,\tau}$ starting from a node in $S_{\sigma,\tau}$ must eventually reach a cycle inside $S_{\sigma,\tau}$. A cycle is a strongly connected component (SCC).  Let $C=\{C1,\ldots ,C_i\}$ be the set of maximal SCCs in $S_{\sigma,\tau}$. Let $C'$ be a directed graph with one node for each SCC in $C$, and an arc from $C_i$ to $C_j$ if there is a path in $S_{\sigma,\tau}$ from $C_i$ to $C_j$.   $C'$ is an acyclic graph with at least one node with no out-arcs.  Let $C_k$ be an SCC in $S_{\sigma,\tau}$ represented by a node with no out-arcs in $C'$. No node in $C_k$ can have any arc at all pointing out of $C_k$, because any path starting with such an arc must eventually cycle inside of $S_{\sigma,\tau}$. By assumption such a cycle cannot be an SCC outside of $C_k$. Suppose there is a path starting with an arc from a node in $C_k$  to a node outside of $C_k$, ending in a cycle contained in $C_k$. The union of $C_k$ and a path leading out of it and back would be an SCC, but this violates the maximality of $C_k$. So there can be no arc out of $C_k$. It follows that $C_k$  satisfies all five criteria in the definition of a bad subgraph. \qed

\pagebreak

\begin{lemma}\label{lemma: stopping game} A simple stochastic game $G$ is a stopping game if and only if it contains no bad subgraphs.
\end{lemma}
Proof: Let $G$ be a simple stochastic game containing a bad subgraph $S$. Let $\sigma$ be a max strategy that chooses arcs pointing into $S$ for all max nodes in $S$, and let $\tau$ be a min strategy that chooses arcs pointing into $S$ for all min arcs in $S$.  Then in $G_{\sigma , \tau}$, there is no path to a terminal from any node in $S$, so $G$ is not a stopping game.  Conversely, let $G$ be a simple stochastic game with no bad subgraphs. By Lemma \ref{lemma: bad subgraph SCC}, $G$ cannot contain any subgraph satisfying criteria 1-4 of the definition of bad subgraphs. Therefore for any subset $V$ of non-terminal nodes of $G$ and any strategy pair $(\sigma, \tau)$, there must exist some node with an arc out of $V$ in $G_{\sigma, \tau}$. It follows that it is possible to construct a path from any node $v_0$ to a terminal in $G_{\sigma,\tau}$, by starting with $V=v_0$ and in each step $i$ finding a node $v_i \in G\setminus V$ such that there is an arc from a node in $V$ to $v_i$, then setting $V=V\cup v_i$. The number of nodes is finite, so eventually a terminal node must be reached. \qed

\subsection{A Simple Stopping Game Generator}

We introduce a simple Stopping Game generator that has a non-zero probability of producing any stopping game with one minor condition. The generator only generates games where no max or min node points directly to a terminal. These nodes can be solved independently of the rest of the graph in constant time (this is proven in Section \ref{sec:badsub}).

The simple Stopping Game generator works in two phases. In phase 1, all nodes are numbered and each non-terminal node is randomly assigned an arc to a higher-numbered node.  Lemma \ref{node numbering} shows that this does not prevent any stopping games from being generated.  In phase 2, all nodes receive second arcs. First, all average nodes receive second arcs uniformly at random.  To assign second arcs to max and min nodes, we pick a max or min node $m$ uniformly at random and find the set $Q$ of all nodes $q$ such that adding arc $(m,q)$ to the graph will not create a bad subgraph. $Q$ is non-empty because there is an always an average node pointing to a terminal, and that node is in $Q$. We then pick a node $v$ from $Q$ uniformly at random and add the arc $(m,v)$ to the graph. We prove below that this is sufficient to guarantee that the final constructed graph contains no bad subgraphs.
\begin{lemma}
\label{node numbering}
Let $G$ be a Stopping Game with n nodes.  Then there is a numbering of nodes in $G$ with the terminals numbered $n-1$ and $n$, such that for every non-terminal node $v$, $v$ has an arc to a higher-numbered node.
\end{lemma}
\emph{Proof.} Fix any strategy pair $(\sigma,\tau)$ in $G$, and let $S = G_{\sigma,\tau}$ be the strategy subgraph induced by $(\sigma,\tau)$. By definition of a Stopping Game, for each node $v$, there is a path in $S$ from $v$ to a terminal.
Let $S^\prime$ be a shortest path tree on $S$ with all paths ending in a terminal node. Define a partial order $P$ on nodes of $S$ as follows:
For any non-terminal node $a$ in $S^\prime$ at distance $k$ from a terminal, let $b$ be the next node in the unique path from $a$ to the terminal (so $b$ is at distance $k-1$ from the terminal).   Let $a<b$ in the partial order $P$. Let $T$ be any total order on the nodes of $S$ that preserves $P$, and number the nodes of $G$ according to $T$.  By construction, every non-terminal node in $G$ has an arc to a higher-numbered node. \qed

\begin{algorithm}[h!t]
    \textbf{node labeling} Number the nodes from $1$ to $n$. Pick integers $a,b,c \geq 1$, such that $n=a+b+c+2$. Label nodes $n-1$ and $n$ as terminal-0 and terminal-1, respectively. Label node $n-2$ as an average node. Assign the remaining numbers to $a-1$ average nodes, $b$ min nodes, and $c$ max nodes uniformly at random.\label{node labeling}\\
    
    \ForEach{non-terminal node $v$}{
        Pick a higher-numbered node $w$ uniformly at random and add an arc from $v$ to $w$.\label{line:first arcs}\\
        \tcc{(Assigns an out-arc to each non-terminal node that has none.)}
    }
    
    \While{there are average nodes with exactly one out-arc}{
        Pick an average node $m$ with exactly one out-arc $(m,p)$ uniformly at random. \\
        Pick any node $q \notin \{m,p\}$ uniformly at random and add arc $(m,q)$. \label{line:avgsecondarc}\\ 
    }
    \While{there are max or min nodes with exactly one out-arc}{
        Pick a node $m$ with exactly one out-arc $(m,p)$ uniformly at random.\label{line:mn}\\
        Let $Q$ be the set of nodes such that arc $(m,q)$ can be added to the graph without adding a bad subgraph, and $q \notin \{m,p\}$ (Algorithm \ref{algo:fastbadsubgraph}).\\
        Remove the terminal nodes from $Q$. \\
       {
            Pick a node $q \in Q$ uniformly at random, and add arc $(m,q)$.\label{line:addmq}\\
        }
        \tcc{(Assigns a second out-arc to each max, and min node.)}
    }

    \Return{The constructed graph.}
    \caption{Stopping Game Generator}
    \label{algo:basicgenerator}
\end{algorithm}

\subsubsection{Proof of Correctness}
\emph{Proof that the generator can produce any stopping game:} By Lemma \ref{node numbering}, in every stopping game the nodes can be numbered in such a way that every non-terminal node has at least one arc to a higher numbered node.  Let $G$ be a stopping game, and let $G^{\prime}$ be a subgraph of $G$ where every non-terminal node has degree 1 and its single arc goes to a higher-numbered node.   Every arc in $G^{\prime}$ can be generated in Line \ref{line:first arcs} of the Stopping Game Generator.  

All remaining arcs can be generated in Lines \ref{line:avgsecondarc}-\ref{line:addmq}. Second arcs from average nodes are generated first, followed by second arcs from max and min nodes, in any order. The only arcs that cannot be added are those that would create a bad subgraph; bad subgraphs persist as additional arcs from max and min nodes are added to the graph, so if a bad subgraph were created at any point, the final constructed graph would not be a stopping game. Consequently, each arc in G can be generated in Lines \ref{line:avgsecondarc}-\ref{line:addmq}.
 
\emph{Proof that the generator produces only stopping games:} Let $G_0$ be the graph constructed in lines \ref{node labeling}-\ref{line:avgsecondarc}. In $G_0$\, each max and min arc has out-degree one. Therefore, there is only one possible pair of max/min strategies, and $G_0$ is also the only possible strategy subgraph.  $G_0$ must be a stopping game, since in $G_0$, every node has a path through successively higher-numbered nodes to a terminal node.  Let $G_k$ be the graph after  $k$ second arcs from max and min nodes have been added, with $k\geq 0$. Assume inductively that $G_k$ is a stopping game, and so it contains no bad subgraphs.  Let $m$ be the next max or min node to receive a second arc. Since Algorithm \ref{algo:fastbadsubgraph} generates a set of arc choices that will not create a bad subgraph, $G_{k+1}$ must also be a stopping game for any of these arc choices.

It remains to be shown that the algorithm for finding valid arcs (Algorithm \ref{algo:fastbadsubgraph}) correctly identifies the set of nodes $Q$ such that $\forall \:\: q \in Q$, adding the arc $(m,q)$ to $G_k$ does not create a bad subgraph. Suppose adding arc $(m,v)$ to $G_k$ would create a bad subgraph $S$. By definition, $S$ is a strongly connected component and all nodes in $S$ must be ancestors of $m$. Algorithm \ref{algo:fastbadsubgraph} removes all ancestors of $m$ from the list of candidates for a new arc, so the initial list $Q$ is guaranteed to contain no node $v$ such that adding arc $(m,v)$ would create a bad subgraph.  Ancestor nodes are then added back to the candidate list one at a time, with a node $q$ restored to the candidate list $Q$ if and only if:
\begin{enumerate} 
\item $q$ is an average node with an arc to a node in $Q$ or a terminal node (Lines \ref{line: initial avg start}-\ref{line: initial avg end}, \ref{line: restore avg start}-\ref{line: restore avg end}); or \label{line: addbackavg}
\item $q$ is a max or min node such that all of its arcs point to a node in $Q$ (Lines \ref{line: restore maxmin start}-\ref{line: restore maxmin end}). \label{line: addbackmaxmin}
\end{enumerate}
Let $S$ be any bad subgraph that could be created from a subset of $m$'s ancestors by adding an arc from $m$ to a node in $S$. Initially, $Q \cap S = \emptyset$. Condition \ref{line: addbackavg} guarantees that any average node restored to the candidate list $Q$ has at least one arc pointing out of $S$ and so could not be contained in $S$. Condition \ref{line: addbackmaxmin} guarantees that any max or min node restored to $Q$ has no arc pointing into $S$, and so could not be contained in $S$. Thus, no arc added back into $Q$ is in $S$, and it remains the case that $Q\cap S=\emptyset$. 

Let $v$ be any node such that adding arc $(m,v)$ would not create a bad subgraph. Suppose $v \notin Q$, the set returned by Algorithm \ref{algo:fastbadsubgraph}.  Let $v$-\textit{reachable} be the set of nodes reachable from $v$ without going through nodes in $Q$. All arcs out of nodes in $v$-\textit{reachable} must either point back into $v$-\textit{reachable} or point into $Q$.  Define the \textit{perimeter} of $v$-\textit{reachable} to be the set of nodes in $v$-\textit{reachable} with at least one arc into $Q$. None of the perimeter nodes can be average nodes, because an average node with an arc into $Q$ would be added to $Q$ in Lines \ref{line: restore avg start}-\ref{line: restore avg end}. Any max or min node in the perimeter must have at least one arc into $v$-\textit{reachable}, or it would have been added to $Q$ in Lines \ref{line: restore maxmin start}-\ref{line: restore maxmin end}. Thus all average nodes in $v$-\textit{reachable} have both arcs pointing into $v$-\textit{reachable} and all max and min nodes in $v$-\textit{reachable} have at least one arc pointing into $v$-\textit{reachable}. By Lemma \ref{lemma: bad subgraph SCC}, $v$-\textit{reachable} must contain a bad subgraph, contradicting the assumption that adding arc $(m,v)$ will not create a bad subgraph. 
Thus all valid arcs in the ancestor set of $m$ must be added back to $Q$.

It follows that the final list $Q$ of valid arcs is correct.
\qed	

\begin{algorithm}[ht!]
    \Input{a  node $m$ that has exactly one out-arc $(m,n)$}
    Let $Q$ be a list containing all of the nodes in the graph.\\
    $Q = Q \backslash \{m, \text{terminal-0},\text{terminal-1}\}$.\\
    
    Let $P = \{m\}$\\
    \While{$P \neq \emptyset$}{
        Let $p$ be a random node $ \in P$.\\
        Set $P = P\backslash \{p\}$.\\
        \ForEach{parent $p^\prime$ of $p$}{
            \If{$p^\prime \in Q$}{
                $Q = Q\backslash \{p^\prime\}$.\\
                Add $p^\prime$ to $P$.\\
            }
        }
    }
     \tcc{(Removes all ancestors from the candidate list.)}
    Let $T = \emptyset$ \\
     \ForEach{node $v \notin Q$}{
        Let $(v,n)$ be $v$'s first out-arc, and $(v,p)$ be $v$'s second out-arc if it exists. \label{line: initial avg start}\\
        
        \If{$v$ is an average node}{
            \If{$n \in \{Q,\text{terminal-0},\text{terminal-1}\}$ OR $p \in \{Q,\text{terminal-0},\text{terminal-1}\}$ if $(v,p)$ exists}{
                Set $T = T \cup \{v\}$.\\
                Set $Q = Q \cup \{v\}$.\\
            }
        }
     } \label{line: initial avg end}
      \tcc{(Finds initial nodes for final processing phase.)}

      \While{$T \neq \emptyset$}{
        Let $p$ be a random node $ \in T$.\\
        Set $T = T\backslash \{p\}$.\\
        \ForEach{parent $p^\prime$ of $p$}{
            Let $(p^\prime,p)$ be $p^\prime$'s first out-arc, and $(p^\prime,u)$ be $p^\prime$'s second out-arc if it exists.\\ \label{line: restore avg start}
            \If{$p^\prime \notin Q$}{
                \If{$p^\prime$ is an average node}
               {
                    Set $T = T \cup \{p^\prime\}$.\\
                    Set $Q = Q \cup \{p^\prime\}$. 
                    \label{line: restore avg end}\\\label{line: restore maxmin start}
                }\ElseIf{$u \in Q$}{
                    Set $T = T \cup \{p^\prime\}$.\\
                    Set $Q = Q \cup \{p^\prime\}$.
                    \label{line: restore maxmin end}\\
                }
            }
        }
      }
      \tcc{(Add back in nodes that don't cause a bad subgraph.)}
    \Return{Q}.
    \caption{Find Valid Arcs}
    \label{algo:fastbadsubgraph}
\end{algorithm}

\FloatBarrier

\section{Simple Reductions}
\label{sec:reductions}
The reduction of SSGs to Stopping Games is not the only useful reduction for generating instances of theoretical interest. In this section we discuss subgraphs that can be solved in polynomial (linear) time, and further reductions. Any research on the complexity of SSGs can safely assume that each instance is a Stopping Game from which all of the following subgraphs have been removed, and to which all of the following reductions have been applied. 

\subsection{Trivially Removable Subgraphs}
\label{sec:badsub}
Let $G$ be a Stopping Game. We show that we may assume for the purposes of complexity analysis that $G$ contains none of the following subgraphs. The strategy will proceed by defining a subgraph $S$ of a Stopping Game $G$, and then showing that a solution to $G$ can be recovered in constant time from a solution to $G\setminus S$.

\begin{description}
  \item [Max or min node with at least one arc to a terminal.] \label{(sub1)} Let $v$ be a max node with arcs $(v,a)$, $(v,b)$ and suppose that $a$ is a terminal node. If $a$ is terminal-0, then $v$ can be merged with $b$ without changing the solution value of any other nodes. If $a$ is terminal-1, then $v$ can be merged with $a$. The reverse is true for min nodes. 
  \item[Any node with two identical arcs.] \label{(sub2)} Let $(v,w)$, $(v,w)$ be the identical arcs. $v$ can be merged with $w$ without changing the solution value of any other nodes.
  \item[An average node $v$ with a self-arc.] \label{(sub3)} Let $v$ have arcs $(v,v)$, $(v,w)$.  Then: \[
    \text{value}(v)= \frac{\text{value}(v) + \text{value}(w)}{2} \Rightarrow \text{value}(v)=\text{value}(w)
    \]
    So, $v$ can be merged with $w$ without changing the solution value of any other nodes.
    \item[A max, min, or average node with in-degree zero.] \label{(sub4)} Let $v$ be a node with in-degree zero. The value of $v$ has no effect on the value of any other node.  We can find a solution to the original SSG by removing $v$, solving the remaining graph, and then solving $v$.
    \item [A terminal with in-degree zero.] \label{(sub5)} Suppose that one of the two terminal nodes has in-degree zero. Then the value of all nodes in the graph is trivially equal to the value of the other terminal node.
\end{description}

After removing the above subgraphs, either $G$ contains at least two distinct average nodes, one with an arc to terminal-1 and one with an arc to terminal-0, or there is only one such average node and the entire graph has a value of $\frac{1}{2}$. In our algorithm, we eliminate the second possibility.

\subsection{Collapsing Clusters}
In this section we provide a linear-time algorithm for solving all nodes with values strictly equal to $1$ or $0$. It is based on the observation that a max node will always choose the arc that points to the higher-value node. There is no higher value than 1, and no lower value than 0. The reverse is true for the min nodes.

The algorithm for finding 1-valued nodes, detailed in Algorithm \ref{algo:collapsingcluster}, starts by defining a cluster set of nodes containing only terminal-0. Any average nodes that point to terminal-0 have a value less than 1 and are added to the cluster. Any min nodes or average nodes that have an arc to the cluster have a value less than 1. Similarly any max nodes with both arcs to the cluster have a value less than 1. Nodes are added to the cluster until no remaining nodes can be added to the cluster. The remaining nodes have a value of 1. The algorithm can be trivially reversed to find 0-valued nodes. 

\begin{algorithm}[h!t]
    Label the nodes from $1$ to $n$, with terminal-0 labeled $n-1$.\\
    Let $q$ be a list containing only terminal-0.\\
    Let $bv$ be a binary string of 1s with length $n$.\\
    Set $bv[n-1]$ to 0. \\

    \While{$q$ is not empty}{
        Let $u$ be a node popped from $q$.\\

        \ForEach{parent $p$ of $u$}{
            Let $i$ be the index of node $p$.\\
            \If{$bv[i]$ is a 1}{
                \If{$p$ is a min or average node}{
                     Set $bv[i]$ to 0.\\
                     Add $p$ to $q$.\\
                }
                Let $a$ and $b$ be the indexes of the nodes that $p$ has arcs to.\\
                \If{$p$ is a max node AND $bv[a] = 0 = bv[b]$}{
                     Set $bv[i]$ to 0.\\
                     Add $p$ to $q$.\\
                }
            }
        }
    }
     \Return{The indexes of the ones in $bv$.}
    \caption{Find 1-Valued Nodes}
    \label{algo:collapsingcluster}
\end{algorithm}

The algorithm is linear because the number of arcs, and therefore parents, is also linear. Each node can be considered only once, and when a node is considered, its parents are iterated over. Each node is a parent to only two nodes and will only appear twice as a parent in the algorithm. Thus, the total number of parents considered is no more than $2n$.

\subsection{Strongly Connected Components}
A strongly connected component (SCC) in a graph is a set of nodes such that every node in the SCC is reachable from every other node in the SCC. Dividing a graph into SCCs can be done in linear time\cite{tarjan1972depth}. Note that the terminal nodes will always be their own SCCs. Once a graph is divided into SCCs, it must be the case that there is an SCC $C$ with no path to any other SCC except for the terminal nodes. The values of the nodes in this SCC are independent of the values of the nodes in the other SCCs. Thus, $C$ can be solved independently from the rest of the nodes. Once it is solved, there must be a new SCC with no path to any other SCC except for the terminal nodes and nodes with constant solved values. The process can be repeated until all of the nodes are solved. In other words, all SCCs can be solved independently, and it is safe to assume that your SSG is a single SCC. However, it is not necessarily true that the complexity of solving a simple stochastic game with multiple SCCs is the same as that of solving a simple stochastic game with only one SCC apart from the terminals. Any individual SCC may have arcs to any number of solved nodes with values $\in (0,1)$. So by assuming that every SSG is a single SCC, you lose the assumption that the only two nodes with constant values are terminal-0 and terminal-1.

\subsection{Useful Assumptions Summary}
\label{subsec:assumptions}
To summarize this section, if you are pursuing a constructive proof that Stochastic Games are in $\mathcal{P}$, you may assume all of the following about the game you are trying to solve:
\begin{enumerate}
    \item It is both a Simple Stochastic Game, and a Stopping Game.
    \item There are no max or min nodes with arcs to the terminal.
    \item There are no nodes with identical arcs or self-arcs.
    \item There are no nodes with in-degree zero.
    \item There is at least one pair of average nodes where one has an arc to terminal-0, and the other has an arc to terminal-1.
    \item There are no nodes with a value of 1 or 0.
    \item \textbf{EITHER} it is a single SCC with the only out-arcs going to nodes with constant values in $[0,1]$, \textbf{OR} there are only two nodes with constant values, terminal-0 and terminal-1.
\end{enumerate}

\subsection{Generator Implementation} \label{section:modifications}

A modified version of the generator algorithm, with all of the modifications from this section (Section \ref{sec:reductions}), is shown on page \pageref{algo:basicgeneratorupdated} as Algorithm \ref{algo:basicgeneratorupdated}. Note that it is possible for this generator to create instances with nodes that have in-degree 0, but it occurs infrequently. When a specific size is desired, such instances can be thrown out and the generator rerun.

\begin{algorithm}[h!t]
    \textbf{node labeling} Number the nodes from $1$ to $n$. Pick integers $a \geq 2$ and $b,c \geq 1$, such that $n=a+b+c+2$. Label nodes $n-1$ and $n$ as terminal-0 and terminal-1, respectively. Label node $n-2$ as an average node, and add an arc from node $n-2$ to terminal-0.  Label node $n-3$ as an average node, and add an arc from node $n-3$ to terminal-1. Randomly assign the remaining numbers to $a-2$ average nodes, $b$ min nodes, and $c$ max nodes.\\
    
    \ForEach{average node $v$ with no out-arcs}{
        Pick a higher-numbered node $w$ uniformly at random and add an arc from $v$ to $w$.\\
        \tcc{(Assigns an out-arc to each non-terminal average node that has none.)}
    }
    
     \ForEach{max and min node $v$}{
        Pick a higher-numbered non-terminal node $w$ uniformly at random and add an arc from $v$ to $w$.\\
        \tcc{(Assigns an out-arc to each max and min node.)}
    }
    Let $z$ be the number of nodes with in-degree zero.\\
    Pick a random number $r$ between $\max(z-(b+c), 0)$ and $\min(a,z)$.\\
    
    \If{$r \neq 0$}{
        \ForEach{integer from $1$ to $r$}{
            Select an average node $m$ with exactly one out-arc $(m,p)$ uniformly at random.\\
            Pick a node $q \notin \{m,p\}$ with in-degree zero uniformly at random, and add arc $(m,q)$. \label{line:avgsecondarc1}\\
        }
    }
    
    \While{there are average nodes with exactly one out-arc}{
        Pick an average node $m$ with exactly one out-arc $(m,p)$ uniformly at random. \\
        Pick any node $q \notin \{m,p\}$ uniformly at random and add arc $(m,q)$. \label{line:avgsecondarc2}\\
    }
    \While{there are max or min nodes with exactly one out-arc}{
        Pick a node $m$ with exactly one out-arc $(m,p)$ uniformly at random.\\
        Let $Q$ be the set of nodes such that arc $(m,q)$ can be added to the graph without creating a bad subgraph, and $q \notin \{m,p\}$ (Algorithm \ref{algo:fastbadsubgraph}).\\

        \eIf{there are nodes with in-degree zero $\in Q$}{
            Randomly pick a node $q \in Q$ that has in-degree zero, and add arc $(m,q)$.\\
        }{
            Randomly pick a node $q \in Q$, and add arc $(m,q)$.\\
        }
        \tcc{(Assigns a second out-arc to each max, and min node.)}
    }
    Merge 1-valued and 0-valued nodes into their respective terminals (Algorithm \ref{algo:collapsingcluster}).\\
    \Return{The constructed graph.}
    \caption{Modified Stopping Game Generator}
    \label{algo:basicgeneratorupdated}
\end{algorithm}

\section{Benchmark Instances}
In this section, we discuss the implementation of our generator algorithm and the benchmark set of problems we generated. The code and benchmark set is publicly available, with documentation explaining how to generate new instances\footnote{https://github.com/IsaacRudich/SimpleStochasticGamesBenchmark}. 

\subsection{Benchmark Set}
We generated instances with sizes in powers of two from $2^5$ to $2^{12}$. For each size, we generated 800 instances, with 100 instances at each ratio of average nodes to max nodes $ \in \{1:4, 2:4, ..., 7:4, 8:4\}$. In order to maintain the intended ratio, some instances are slightly larger or slightly smaller than the labeled size. For example, instances labeled with size $4096$ and ratio $1:4$ have $1820$ max nodes, $1820$ min nodes, $455$ average nodes, and $2$ terminal nodes, for a true total of $4097$ nodes.

Each instance in our benchmark set is fully reduced using all of the reductions listed in Section \ref{subsec:assumptions}, including that each instance is a single SCC plus two terminal nodes. This was accomplished by simply generating instances until we found enough in each category that were already fully reduced when they were generated. This means that each instance in our benchmark set has the property of being both a single SCC and having only two nodes with constant value. However, if you are interested in studying the behavior of instances with several constant-valued nodes, that can be easily accomplished using the same instances. Simply reassign each arc that points to a terminal to a node with a random fixed value $\in [0,1]$. The instance will still be fully reduced, and will still be a single SCC.

\subsection{Experimental Results}
We solved each of the $6400$ problems in our benchmark set $100$ times using random seeds with two different algorithms and recorded the number of iterations and time to solve for each run. The experiments were performed on a computer equipped with an 2023 M2 Pro and 32Gb RAM. The first algorithm tested was Hoffman-Karp, the standard value-iteration algorithm. The second algorithm tested was an implementation of a permutation improvement algorithm based on an algorithm from Gimbert and Horn \cite{permutation_improvement}. We achieved a substantial speed-up over a naive implementation of the permutation improvement algorithm by generalizing the concept of collapsing clusters to nodes of any value, then pre-calculating the collapsing clusters that would be implied by any given candidate ordering of the average nodes. 

\begin{figure}[!ht]
    \centering
     \includegraphics[scale=.08]{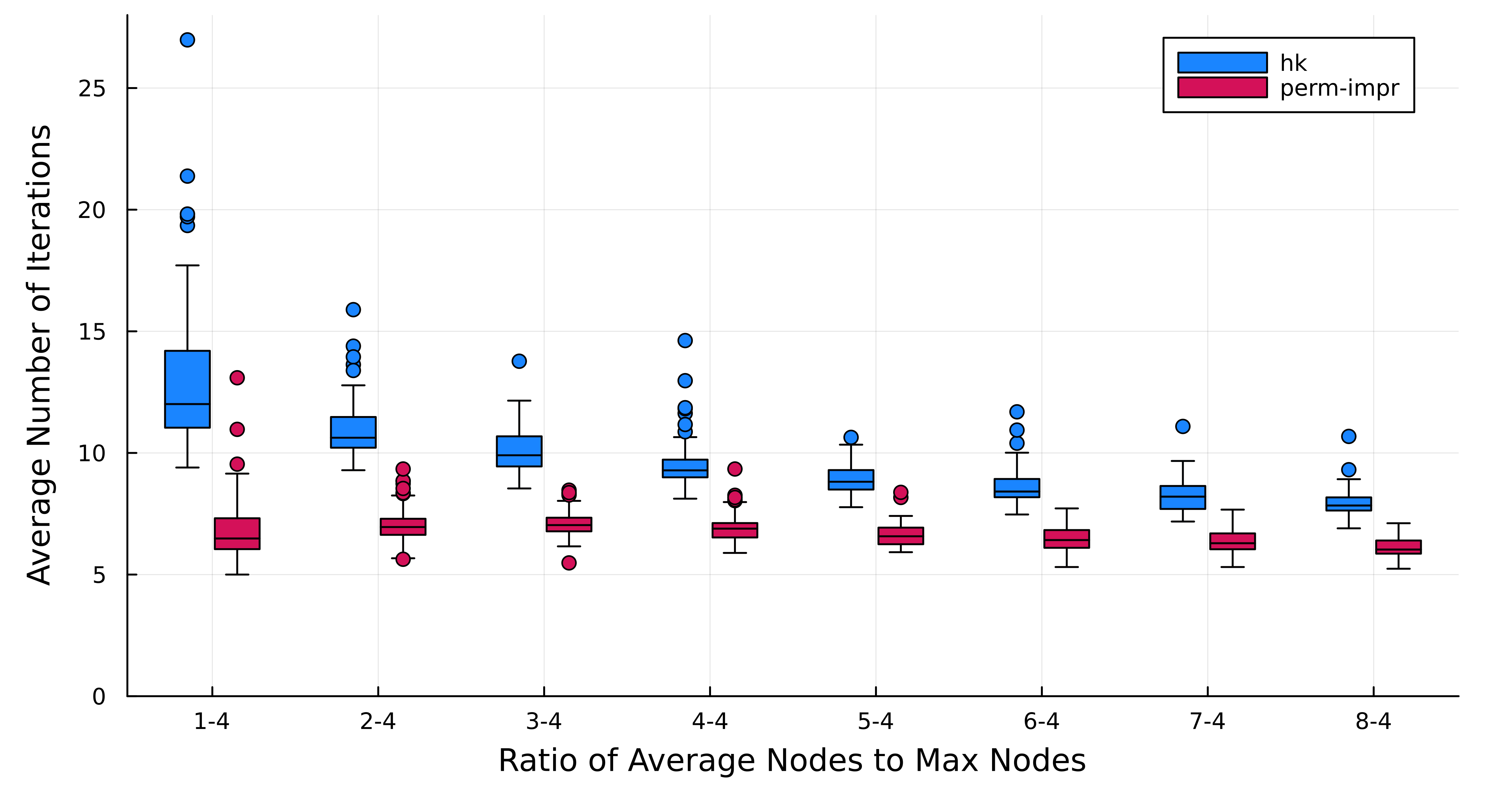}
    \label{fig:4096graph}
    \caption{Average Iterations to Solve for Size 4096 Games}
\end{figure}

Figure \ref{fig:4096graph} shows data for the two algorithms, grouped by ratio, for the size $4096$ SSGs. The graphs for the other sizes look nearly identical but with globally lower numbers of iterations; they are available with the code for the generator. The data shows that the permutation improvement algorithm consistently outperforms Hoffman-Karp in terms of both iterations and time. Hoffman-Karp performs better as the number of decision nodes goes down, while the permutation improvement algorithm shows consistent performance for all of the ratios tested. 

Additional data showing the average run-time, in both time and iterations, for both algorithms and each category of problem is available in Appendix \ref{app:data} Tables \ref{tab:avgitr}, \ref{tab:avgitr_m}, \ref{tab:avgtime}, and \ref{tab:avgtime_m}. In terms of average number of iterations, the permutation improvement algorithm outperformed Hoffman-Karp on every category of problem. In terms of average amount of time, the permutation improvement algorithm outperformed Hoffman-Karp on almost every category of problem. The 4 categories (out of 48) where Hoffman-Karp out-performed are shown in bold.

\section{Conclusion}
This paper presented a fast algorithm for generating Simple Stochastic Stopping Games. It also provided several polynomial reductions and assumptions for further simplifying and reducing Simple Stochastic Games. These assumptions are useful for anyone attempting a constructive proof that the complexity of Stochastic Games is in $\mathcal{P}$. They are also useful for trying to understand why hard-to-solve instances are so challenging to generate. Our code for generating instances is open source, as well as the the benchmark instances we generated. 

\bibliography{sg} 
\bibliographystyle{ieeetr}

\appendix
\section{Experimental Data}
\label{app:data}
\begin{table}[ht!]
    \centering
\begin{tabular}{c|c|c|c|c|c|c}
\hline
\tiny{\diagbox{ratio}{size}} & 128 & 256 & 512 & 1024 & 2048 & 4096 \\
\hline
1-4 & 5.5 & 7.2 & 8.6 & 9.9 & 11.4 & 12.9 \\
2-4 & 5.6 & 7.0 & 8.2 & 9.2 & 10.1 & 10.9 \\
3-4 & 5.4 & 6.5 & 7.6 & 8.5 & 9.5 & 10.1 \\
4-4 & 5.3 & 6.2 & 7.0 & 8.0 & 8.8 & 9.5 \\
5-4 & 4.9 & 5.9 & 6.7 & 7.4 & 8.3 & 8.9 \\
6-4 & 4.8 & 5.6 & 6.4 & 7.2 & 7.8 & 8.6 \\
7-4 & 4.6 & 5.3 & 6.0 & 6.8 & 7.6 & 8.3 \\
8-4 & 4.3 & 5.2 & 5.9 & 6.5 & 7.3 & 7.9 \\
\end{tabular}
    \caption{Average Hoffman-Karp Iterations}
    \label{tab:avgitr}
\end{table}

\vspace{5mm}

\begin{table}[ht!]
    \centering
    \begin{tabular}{c|c|c|c|c|c|c}
\hline
\tiny{\diagbox{ratio}{size}} & 128 & 256 & 512 & 1024 & 2048 & 4096 \\
\hline
1-4 & 2.2 & 3.0 & 3.9 & 4.8 & 5.9 & 6.8 \\
2-4 & 2.9 & 3.8 & 4.9 & 5.8 & 6.4 & 7.0 \\
3-4 & 3.3 & 4.1 & 5.0 & 5.9 & 6.4 & 7.1 \\
4-4 & 3.4 & 4.2 & 4.9 & 5.6 & 6.3 & 6.9 \\
5-4 & 3.4 & 4.1 & 4.9 & 5.4 & 6.1 & 6.6 \\
6-4 & 3.4 & 4.0 & 4.8 & 5.4 & 5.9 & 6.5 \\
7-4 & 3.3 & 4.0 & 4.6 & 5.2 & 5.8 & 6.4 \\
8-4 & 3.2 & 3.9 & 4.4 & 5.1 & 5.6 & 6.1 \\
\end{tabular}
    \caption{Average Permutation Improvement Iterations}
    \label{tab:avgitr_m}
\end{table}

\pagebreak

\begin{table}[ht!]
    \centering
\begin{tabular}{c|c|c|c|c|c|c}
\hline
\tiny{\diagbox{ratio}{size}} & 128 & 256 & 512 & 1024 & 2048 & 4096 \\
\hline
1-4 & 5.0 & 12.1 & 29.1 & 71.1 & 194.2 & 1068.4 \\
2-4 & 5.0 & 12.3 & 28.6 & 74.6 & 198.3 & 1497.1 \\
3-4 & 5.0 & 11.7 & 27.3 & 71.4 & 203.2 & 2769.9 \\
4-4 & 4.8 & 11.1 & 25.6 & 69.5 & 199.6 & 4096.5 \\
5-4 & 4.4 & 10.5 & 24.7 & 64.7 & 195.0 & 3934.0 \\
6-4 & 4.2 & 10.1 & 23.7 & 64.1 & 187.2 & 4308.5 \\
7-4 & 4.1 & 9.4 & \textbf{22.2} & \textbf{59.9} & 185.2 & 4378.9 \\
8-4 & 3.8 & 9.2 & \textbf{21.6} & \textbf{58.4} & 181.2 & 5330.0 \\
\end{tabular}
    \caption{Average Time in Milliseconds for Hoffman-Karp (\textbf{bold} indicates that HK outperformed perm-impr)}
    \label{tab:avgtime}
\end{table}

\vspace{5mm}

\begin{table}[ht!]
    \centering
\begin{tabular}{c|c|c|c|c|c|c}
\hline
\tiny{\diagbox{ratio}{size}} & 128 & 256 & 512 & 1024 & 2048 & 4096 \\
\hline
1-4 & 1.9 & 4.9 & 13.1 & 34.4 & 103.2 & 591.5 \\
2-4 & 2.6 & 7.0 & 18.3 & 50.6 & 135.8 & 945.8 \\
3-4 & 3.1 & 8.1 & 20.4 & 56.1 & 157.9 & 1817.7 \\
4-4 & 3.3 & 8.6 & 21.3 & 59.1 & 168.0 & 3210.5 \\
5-4 & 3.4 & 8.7 & 22.8 & 58.7 & 174.0 & 3326.0 \\
6-4 & 3.5 & 8.9 & 22.7 & 62.6 & 177.8 & 3195.3 \\
7-4 & 3.5 & 9.1 & \textbf{22.7} & \textbf{62.8} & 183.3 & 3922.9 \\
8-4 & 3.6 & 9.2 & \textbf{22.5} & \textbf{62.8} & 180.8 & 3918.1 \\
\end{tabular}
    \caption{Average Time in Milliseconds for Permutation Improvement (\textbf{bold} indicates that HK outperformed perm-impr)}
    \label{tab:avgtime_m}
\end{table}

\end{document}